\documentclass[11pt]{article}
\usepackage{graphicx}
\begin{document}

\begin{titlepage}

\begin{center}

{\bf \Large Type Ia supernovae in dense circumstellar gas}

\bigskip

N.N. Chugai, L.R. Yungelson

\bigskip

Institute of Astronomy, RAS\\
Moscow 119017, Pyatnitskaya 48\\

\end{center}

\end{titlepage}

\newpage

We propose
 a simple model for the bolometric light curve of
type Ia supernova exploding in a dense  circumstellar (CS)
envelope 
to describe the light curves of supernovae 2002ic and 1997cy.
The modeling shows that at the radius
$\sim7\times10^{15}$ cm the density of CS envelopes
 around both supernovae is similar.
The mass of the CS envelope around SN~1997cy is 
close to $5~M_{\odot}$, while the characteristic time of the 
ejection of this envelope does not exceed 600 yr.
We analyze two possible evolutionary 
scenarios which might lead
to the explosion of type Ia supernova inside a
 dense CS envelope:
accretion on CO white dwarf in the symbiotic binary and
evolution of a single star with the initial mass of about $8~M_{\odot}$.
If the conjecture about the explosion of type Ia supernova in
a dense CS envelope is correct in the case of
SN~2002ic and SN~1997cy then
the rapid loss of the red supergiant envelope
and the subsequent explosion of the CO white dwarf
are synchronized by certain mechanism.
This mechanism might be related to the contraction of the
white dwarf as it approaches the Chandrasekhar limit.
 We show that formation of a (super)Chandrasekhar mass
 due to the merger of a CO
 white dwarf and CO core of a red supergiant 
 with consequent explosion is unlikely, since it does not 
 provide the required synchronization  
 of the rapid mass loss and explosion.

\newpage

\section{Introduction}

Recently Hamuy et al. (2003) upon the basis of the
spectra and light curve of the supernova
SN~2002ic with narrow H$\alpha$ emission suggested
that this object was a SN~Ia interacting with a dense 
circumstellar (CS) envelope.
Furthermore, Hamuy et al. claim that another supernova,
SN~1997cy, classified as SN~IIn (i.e., the supernova with narrow
H$\alpha$ emission) is a counterpart of SN~2002ic.
The narrow unresolved H$\alpha$ line
is attributed to the emission of the 
photoionized dense CS gas, while  H$\alpha$ broad component with the
full width at half maximum of $\approx 1800$ km s$^{-1}$
is identified with the emission of shocked dense CS clouds.
The dense CS matter (CSM) according to Hamuy et al. (2003) 
is associated with
a high mass-loss from the red supergiant. The latter
may be either a companion to a white dwarf in  
a symbiotic binary system, or
a presupernova itself, if supernova originates from
a single intermediate mass star in the
SN\,1.5 scenario (Iben and Renzini 1983).

The luminosity of SN~2002ic is high by SN~Ia standards
and this is attributed
to the interaction with the dense CS matter (Hamuy et al. 2003).
Remarkably, the light curve of
SN~1997cy has been already modeled in terms of the CS interaction
(Turatto et al. 2000). However, in the latter study
SN~1997cy was claimed to be a hypernova with 
the enormous energy of $3\times10^{52}$ erg and large
mass of $25~M_{\odot}$. 
The hypernova model is, of course, incompatible with
the hypothesis of SN\,Ia as it was admitted by Hamuy et al. (2003).

On the other hand, it was demonstrated recently that the hypernova model
is not necessary at all to account for
the bolometric light curve of SN~1997cy;
the CS interaction of a normal supernova with a typical energy
of $10^{51}$ erg and low mass of $1.5~M_{\odot}$ is quite successful in
reproducing the light curve of SN~1997cy (Chugai and Danziger 2003).
This result along with the spectral arguments in favor of
explosion of SN~Ia in the case of SN~1997cy
(Hamuy et al. 2003) compels us to consider the explosion of
 SN~Ia inside a dense CS envelope as a promising  
 interpretation of SN~2002ic and SN~1997cy.

Despite it is not clear as yet how the spectrum of
SN~Ia forms in the case of the strong CS interaction,
it would be sensible to study
 the density and the structure of the CS envelope using
constraints imposed by the bolometric light curve.
Here we model the light curves of both aforementioned
supernovae assuming that
their radiation is a combination of the radioactive
luminosity of SN~Ia and the luminosity powered by the
interaction of the supernova with the dense CS environment.
The 
results are discussed then in terms
of different evolutionary scenarios presumably leading
to SN~2002ic and SN~1997cy events.

\section{The light curve model}

Let us assume that SN~Ia explodes inside a spherically-symmetric
CS envelope. We are interested in the bolometric light curve
produced by the superposition of the
intrinsic radioactive luminosity of SN~Ia and of the luminosity
powered by the interaction with the CS matter.
To compute the light curve we use the model which was
applied for the modeling of the light curve of SN~1998S (Chugai 2001).
In this model the interaction dynamics is calculated in
the thin shell approximation which treats the region between
the forward and reverse shock waves as infinitely thin shell
(Chevalier 1982).
The supernova envelope is characterized by the mass ($M$), kinetic
energy ($E$), and presupernova radius ($R_0$). We assume that
the initial kinematics of the supernova envelope is
 free expansion ($v\propto r$) and the
density distribution in this envelope is exponential, 
$\rho\propto \exp\,(-v/v_0)$, where $v_0$ is defined by 
$E$ and $M$.
Below we adopt $M=1.4~M_{\odot}$,
$E=1.5\times10^{51}$ erg, and the $^{56}$Ni mass of
$0.7~M_{\odot}$. This choice is consistent with SN~Ia delayed detonation
models with maximal  $^{56}$Ni mass (H\"{o}flich et al. 1995).

The numerical solution of the equation of motion for the thin shell 
provides us with the evolution of the shell radius and velocity.
The resulting kinetic luminosities of the forward and
reverse shock waves are transformed into X-ray
luminosities using cooling rates of the post-shock gas (Chugai 1992).
The X-ray radiation of both shocks with corresponding temperatures 
is partially absorbed by the supernova envelope,
thin shell and the CS gas. The bolometric luminosity 
powered by the interaction we calculate as the total 
absorbed X-ray luminosity.
This simple model ignores details of the optical spectrum formation 
and cannot describe the emergent spectrum in detail.

The light curve powered by the radioactive decay
$^{56}$Ni---$^{56}$Co---$^{56}$Fe
is calculated using analytical theory for homogeneous 
envelope (Arnett 1980).
To compute the overall energy deposition rate
we assume that $^{56}$Ni is mixed in the inner $0.8~M_{\odot}$.
The interaction of gamma-rays with the matter is treated in
the single flight approximation assuming the absorption coefficient
of 0.03 cm$^2$ g$^{-1}$. The opacity is assumed to be
0.15 cm$^2$ g$^{-1}$.

The calculated bolometric curves of SN~Ia without 
and with the CS interaction assuming
 the wind with a constant
mass loss rate ($\rho\propto r^{-2}$) are shown in Fig.~1.
The density parameter $w=4\pi r^2 \rho$ for the 
 models in Fig.~1
is  $4\times 10^{15}$ g cm$^{-1}$,
$2\times 10^{16}$ g cm$^{-1}$, and $10^{17}$ g cm$^{-1}$, respectively.
For the wind velocity $10$ km s$^{-1}$ these values correspond
to the mass loss rate $6\times10^{-5}~M_{\odot}$~yr$^{-1}$,
 $3\times10^{-4}~M_{\odot}$~yr$^{-1}$, and
$1.5\times10^{-3}~M_{\odot}$~yr$^{-1}$.
Figure 1 shows that the contribution of the interaction 
to the
SN~Ia luminosity at the light maximum becomes substantial for
$w>10^{16}$ g cm$^{-1}$, while 
the contribution of the interaction with
the more rarefied wind, $w\sim10^{16}$ g cm$^{-1}$ can be detected 
only at the very late epoch ($t> 1$ yr).
Amazingly, at late time ($t\sim 300$ d), the
 interaction luminosity for low CS density 
depends on $w$ more steeply than for high CS density, i.e., 
there is a saturation effect.
This is related to the fact that for low $w$ 
the luminosity is determined primarily by the radiative reverse shock
wave while the contribution of the adiabatic forward shock 
increases with $w$. For high $w$ the forward shock wave
becomes radiative and dominant, so the overall luminosity 
turns out to be
$L\propto wv_{\rm s}^3$, (where $v_{\rm s}$ is the
thin shell velocity). In this case the luminosity obviously 
 grows less rapidly than $w$
since the velocity $v_{\rm s}$ notably decreases as $w$ increases.

We considered above only the case of the stationary wind. However,  
the shape of the light curve powered by the CS interaction
depends on the density distribution in the CS envelope.
This fact will be taken into account below in the modeling
of the light curves of SN~2002ic and SN~1997cy.

\section{Light curves of SN~2002ic and SN~1997cy and their
CS envelopes}

The $BVI$ light curves during 70 day interval
are available for SN~2002ic.
To recover monochromatic light curves using our bolometric light curve
model one needs to specify the photosphere radius
($R_{\rm p}$). We assume that the photosphere coincides with the
thin shell, i.e, $R_{\rm p}=R_{\rm s}$. This equality is
justified for SN~1998S (SN~IIn) where  the opaque
thin shell forms at the SN/CSM interface (Chugai 2001).
However, for SN~2002ic this assumption should be considered as a rough
approximation since the spectrum does not show the smooth continuum which 
is characteristic of SN~1998S.
Adopting $R_{\rm p}=R_{\rm s}$ and using the calculated
bolometric luminosity we derive the effective temperature and
assuming black-body spectrum we then compute absolute $BVI$ magnitudes.
To compare the model with observations we adopt redshift $z=0.067$,
reddening $E(B-V)=0.073$ (Hamuy et al. 2003) and the Hubble constant
$H_0=70$ km s$^{-1}$ Mpc$^{-1}$. The $K$-correction related to
the redshift is small and was calculated assuming black-body spectrum
with the temperature $T=10000$ K.
We adopt the supernova explosion date 2452585~JD.

The $BVI$ light curves for SN~2002ic are calculated for a
wide range of parameter variations.
The general conclusion is that the stationary wind
$\rho\propto r^{-2}$ (model ic2) results in the steep light
drop which is inconsistent with observations.
More adequate behavior is provided by the model ic1 with the
flat density distribution (Fig. 2).
The light curves for both models are shown in Fig.~3.
Although, we focus on the modeling of $V$ light curve, which 
possesses pre-maximum observational points, in the other
 bands the agreement is also satisfactory.
In  the $I$ band the accuracy of the fit for the model
ic1 is better than 0.2 mag.
 Deviations are greatest in the $B$
band, although the general behavior is reproduced quite sensibly.
The bolometric light curves for both models are shown in
Fig. 4. It demonstrates that CS interaction contributes 
about 50\% at the light maximum and dominates after the maximum.
 This inference is qualitatively consistent with 
 the decomposition of 
 the light curve of SN~2002ic by Hamuy et al. (2003).
Note, for the interaction to be prominent at the light maximum,
the CS envelope must be dense at the radius $r\sim2\times10^{15}$ cm.
With the expansion velocity $u\sim10$ km s$^{-1}$ the age of 
this matter is $\sim r/u\sim70$ yr; i.e., the supernova 
must explode not later than $\sim10^2$ yr after the 
termination of the major mass loss episode. 

Some uncertainty in the choice of the photospheric radius
may affect the derived CS density. To check the effect of 
this uncertainty we calculated monochromatic
light curves of SN~2002ic assuming $R_{\rm p}=0.8R_{\rm s}$.
The best fit is obtained for the CS density 10\% larger than that
for the model ic1. This demonstrates that the derived CS density
is not very sensitive to the assumption on the 
photospheric radius.

The lack of photometric data for SN~2002ic at late epoch 
($>70$ days) does not permit us to derive the CS density at
distances exceeding $\approx7\times10^{15}$ cm. 
Within this radius, the integrated mass of the CS envelope is 
$\approx 0.4~M_{\odot}$.

Fortunately, for SN~1997cy the light curve is traced for over  
600 days after the outburst (Turatto et al. 2000).
The bolometric light curve of SN~1997cy 
published by Turatto et al. (corrected for the adopted by us 
$H_0=70$ km s$^{-1}$ Mpc$^{-1}$) is shown in Fig.~5.
In this plot we also present our simulated
light curves for models cy3 and cy4 with CS density profiles
 shown in Fig.~6.
The model cy3 has the same density distribution as the model 
ic1 for SN~2002ic, with the exception that the density is 10\% higher 
in the cy3 model.
The model cy4 with the inner density minimum shows the similar fit
of the light curve as the model cy3 and thus demonstrates the uncertainty 
in the choice of the CS density distribution in the inner region. 
Yet we note that the CS densities around SN~1997cy and SN~2002ic 
are roughly similar  
at the radii of $\sim 7\times10^{15}$ cm. 

The steepening of the light curve after about day 500 indicates
the CS density drop at the radius
$r=R_{\rm b}\approx 2\times10^{16}$ см (Figs.~5 and 6). The integrated
mass of the CS envelope is
$M_{\rm cs}=5.4~M_{\odot}$ for the model cy3 and
$5.9~M_{\odot}$ for the model cy4. We adopt the "minimal" 
 estimate $M_{\rm cs}=5.4~M_{\odot}$.
 The recovered CS mass 
for SN~1997cy is close to that found earlier in the low mass supernova
model (Chugai and Danziger 2003). More surprising is that 
the mass of the CS envelope recovered  by Turatto et al. (2000)
in their model of hypernova is also practically the same 
($\approx5~M_{\odot}$).

The age of the dense CS envelope around SN~1997cy
is $t_{\rm cs}=R_{\rm b}/u=630/u_{1}$ yr, where $u_{1}$
is the flow velocity in units of 10 km s$^{-1}$.
The average mass loss rate is, therefore,
\begin{equation}
\dot{M}=M_{\rm cs}/t_{\rm cs}=8\times10^{-3}\left(\frac{M_{\rm cs}}
{5~M_{\odot}}\right)u_1 
\quad M_{\odot}~\mbox{yr}^{-1}.
\end{equation}
The velocity of the superwind
is generally $10-20$ km s$^{-1}$  (Wood 1993), so the average
mass loss rate for the presupernova of
SN~1997cy assuming CS mass $5.4~M_{\odot}$
is $\dot{M}\sim10^{-2}~M_{\odot}$~yr$^{-1}$.
Remarkably, given the similar density of the CS envelope around 
SN~2002ic, the estimated average mass loss rate is characteristic 
of this event too.

Let us now check whether the derived CS density distribution around
SN~1997cy is consistent with the luminosity of
the narrow H$\alpha$ emission.
The emission measure of the CS envelope in the model cy3
on day 70 is $6.4\times10^{65}x^2$ cm$^{-3}$ (where $x$ is
the hydrogen ionization degree).
Adopting the electron temperature $10^4$\,K, the predicted 
recombination luminosity of H$\alpha$ in Menzel case B is
$\approx 2\times 10^{41}x^2$ erg s$^{-1}$.
Turatto et al. (2000) distinguish in the H$\alpha$ profile
on day 70 three components --- narrow, intermediate, and broad.
However, the relative contribution of each component is
not reported. To minimize uncertainty related to the
broad component which is affected by the blend of Fe II lines,
we consider only the intermediate component and the narrow one
within the range of radial velocities $|v_r|<2000$ km s$^{-1}$.
The latter is the range of the velocities of the broad 
H$\alpha$ component in SN~2002ic either (Hamuy et al. 2003).
The spectrum of SN~1997cy on day 70 (Turatto et al. 2000)
provides a rough luminosity estimate for these two components
$L\sim 5\times10^{40}$ erg s$^{-1}$. 
According to the data on SN~2002ic obtained by Hamuy et al. (2003) we assume
that
the contributions of narrow and intermediate (i.e., broad in SN~2002ic)
components are
comparable. With this assumption the luminosity of the 
narrow component in SN~1997cy is 
$\sim 2.5\times10^{40}$ erg s$^{-1}$. Combined with 
the luminosity suggested by the emission measure, the latter 
luminosity implies the average ionization degree
in the CS envelope $x\sim 0.35$. This suggests the Thomson 
optical depth of the CS envelope of $\tau_{\rm T}\approx 0.5$.

Remarkably, the luminosity of narrow H$\alpha$ in SN~2002ic
during whole observed period ($t\leq 70$ days) is practically
constant and equal to $\approx 2\times10^{40}$ erg s$^{-1}$
(Hamuy et al. 2003). We thus conclude that the 
 luminosity of narrow H$\alpha$
in both supernovae is similar at the epoch of 70 days.
This fact taken together with 
the similarity of the CS density at $r\sim7\times10^{15}$ cm
indicates that for SN~2002ic
the structure of the CS envelope 
and its total mass are roughly similar to those of SN~1997cy. 
We thus expect that the mass of the CS envelope of SN~2002ic is 
probably also several solar mass.

\section{The nature of presupernova}
\label{sec:presn}

The striking feature of progenitors of 
both SN~2002ic and SN~1997cy is an unusually high mass loss rate
 $\dot{M}\sim 10^{-2}~M_{\odot}$ yr$^{-1}$ during 
 several hundred years prior to the explosion.
 This indicates the existence of certain 
mechanism that synchronizes 
 the phase of violent mass loss with the epoch of supernova explosion. 
 If  SN~2002ic and SN~1997cy actually are SNe\,Ia, 
 it is naturally to assume that the violent mass loss is 
 related to the epoch when CO white \mbox{dwarf} attains the Chandrasekhar 
 mass ($M_{\rm Ch}$).

Following the hypothesis of Hamuy et al. (2003) we consider 
two sce\-na\-rios that may, in principle, result in an explosion of a 
 SNe\,Ia inside hydrogen-rich circumstellar envelope: 
 (i) an explosion of accreting white dwarf in a binary system 
 with a red supergiant companion to the white dwarf 
 (Whelan and Iben 1973) and (ii) an explosion of a degenerate 
  CO core of a single star 
 (Arnett 1969; Iben and Renzini 1983).

\subsection{Binary system}
\label{subsec:binary}

The model of binary system suggests that the CO white dwarf 
attains the Chandrasekhar mass as a result of accretion 
from a red 
giant or supergiant companion. Since the red supergiant 
mass loss rate through the superwind does not exceed 
$3\times10^{-4}~M_{\odot}$~yr$^{-1}$ (Wood 1994),
rather unusual conditions are necessary for the mass loss rate to 
attain $10^{-2}~M_{\odot}$~yr$^{-1}$.
The required mass loss rate, as well as synchronization of explosion 
with the stage of intense mass loss, could be qualitatively 
explained within two feasible sequences of events in a binary.

The first version exploits the steepening of 
the mass-radius relation for white dwarfs when approaching 
Chandrasekhar limit. For $M\approx M_{\rm Ch}$ 
a \mbox{small} increase of dwarf mass is accompanied by a considerable 
reduction of its radius. Even more, it may happen that as a result 
of nuclear burning in the vicinity of the maximum of temperature,  
a considerable fraction of the outer layers of CO dwarf may convert 
into  $^{22}$Ne; this also may lead to some reduction of the radius
 of the dwarf in several hundred years immediately before the explosion. 
The decrease of white dwarf radius will result in the spin-up of its 
 rotation. The rapid rotation and enhancement of magnetic field due to 
the differential rotation combined with the increase of the 
gravitational
 potential at the surface of white dwarf will result in the increase 
 of the kinetic luminosity of the wind from the innermost parts of accretion 
 disk. 
We assume that, similarly 
 to the winds from white dwarfs (Hachisu et al. 1999), the interaction
 of this wind with the envelope of the red 
  supergiant may  cause an intense mass loss by the latter. This 
  effect was considered by Hachisu et al. (1999) as a 
  mechanism for the removal of angular momentum.  We are 
  interested here in the supergiant mass loss only.    

Let the mass and radius of white dwarf be $M_1$ and $R_1$, 
while those of red supergiant be $M_2$ and $R_2$. Assuming 
that the velocity of white dwarf stellar wind is of the order 
of the escape velocity (the maximum estimate) and the velocity 
of gas lost by red supergiant is  of the order of  escape
velocity as well, one gets, from the condition of the balance of 
energy fluxes (the regime of ablation), an estimate for the typical 
mass loss rate by the red supergiant:
\begin{equation}
\dot{m}_2= f(q)\dot{m}_1 
\left(\frac{M_1}{M_2}\right)\left(\frac{R_2}{R_1}\right)\,,
\label{eq:mdot2}
\end{equation} 
where $\dot{m}_1$ is the mass loss rate via the 
fast wind of white 
dwarf, $f(q)$ is the geometrical factor calculated by 
Hachisu et al. (1999), and $q=M_2/M_1$.
Given the estimated mass of the CS envelope we adopt $q\approx4$. 
For $f(q)$ we take $f\approx0.033$ assuming that the red giant fills its 
Roche lobe (Hachisu et al. 1999).
Inserting these values in Eq. (\ref{eq:mdot2})
and adopting  $R_2=1000~R_{\odot}$, $R_1=3\times10^{-3}~R_{\odot}$, 
and $\dot{m}_1=10^{-5}~M_{\odot}$ yr$^{-1}$ (this value corresponds to 
the white dwarf accretion rate limit set by Eddington luminosity) 
we obtain $\dot{m}_2\approx 10^{-2}~M_{\odot}$ yr$^{-1}$.
The model, thus, is able to provide the required red-giant mass loss rate.
However, it should be noted that this value 
is obtained for rather extreme assumptions about white dwarf mass 
loss rate and wind velocity, and assuming  
100\% efficiency of the ablation mechanism. 

A modification of the scenario of the rapid mass loss 
in the symbiotic binary with 
the white dwarf mass approaching $M_{\rm Ch}$ 
might be the formation of a common envelope. This might happen 
due to the expansion of the red supergiant envelope induced by the white 
dwarf wind and subsequent Roche-lobe overflow. 
The common envelope also could result in
the loss of the envelope by the red supergiant in the time scale of 
several hundred years.

The drawback of mechanisms of the 
rapid mass loss induced by the white dwarf 
contraction at the mass $M\approx M_{\rm Ch}$ 
is the relatively weak dependence of the kinetic luminosity of the 
white dwarf wind on the dwarf radius.
This leaves us with a troublesome question as to why 
the envelope is lost at the right time and not markedly earlier.

The second version of the symbiotic star evolution suggests 
the formation and loss of the common envelope with the subsequent merger 
of the sub-Chandrasekhar CO white dwarf and the degenerate CO core 
of the red supergiant. 
The formation of a super-Chandrasekhar mass object followed by the 
SN~Ia explosion would explain then synchronization of the violent loss 
of the hydrogen envelope and SN\,Ia event.
However, this prima facie natural 
synchronization mechanism suffers from the following serious problem. 

For the merger of dwarf and core due to the gravitational wave radiation 
to occur soon after the termination of the common
 envelope stage, the dwarf -- core pair has to become rather close. 
 For the merger to occur in $t_0$ years, the initial semi-major axis of the 
 dwarf -- core system should be (Landau and Lifshitz 1971)
 
\begin{equation}
a=2\times10^9\left(\frac{t_0}{100\;\mbox{\rm yr}}\right)^{0.25}
 \left(\frac{M}{M_{\odot}}\right)^{0.5}
\left(\frac{\mu}{M_{\odot}}\right)^{0.25}\;\mbox{cm},
\label{eq:a0}
\end{equation}
where $M$ is the total mass of the system and  $\mu$ 
is its reduced mass.
 For instance, for similar masses of components of $0.8~M_{\odot}$ and 
$t_0=100$ yr the distance between components after the common envelope 
stage has to be  
$a=2\times10^9$ cm. Such a close approach is accompanied by 
the release of 
$8.6\times10^{49}$\,erg of the binding energy with $4.3\times10^{49}$\,erg 
deposited into envelope in the form of hydrodynamic motions.
This estimate only weakly depends 
on the masses of the dwarf and the core. If all this energy 
is spent on the ejection of the envelope, then, given 
the low binding energy of the red giant envelope  ($<10^{48}$ erg), 
 the kinetic energy of the circumstellar envelope 
 should be $\sim4\times10^{49}$ erg.
  In reality, for the mass of the CS envelope of 
  $5~M_{\odot}$ and velocity of its expansion $<300$ km s$^{-1}$ 
  (Hamuy et al. 2003), the kinetic energy of the envelope is only
  $<5\times10^{48}$\,erg, i. e., at least by an order of 
  magnitude lower. Thus, the energy released during the spiral-in
  cannot be  
  spent entirely on the ejection of the envelope. On the other hand, 
  such a huge energy cannot be radiated away in several hundred 
  years as well. Actually, the maximum average luminosity of 
  a gravitationally 
  bound red supergiant with initial mass $<10\,M_{\odot}$ 
does not exceed $10^5\,L_{\odot}$ (Iben and Renzini 1983).  
Even in the case of maximum luminosity the total energy 
radiated away in 600 yr does not exceed $8\times10^{48}$ erg.
The rate of energy generation during the spiral-in that exceeds the  
rate of radiation by the 
hydrostatic configuration, would result in a rapid expansion 
of the envelope and in turn-off of the spiral-in process. 

To summarize, in the stage of intense mass loss 
from the common envelope the dwarf and the core cannot get 
closer that it is allowed by the energy loss via
 radiation and mass 
loss ($\sim 10^{49}$ erg). In this case the minimum final 
separation of the objects after common envelope phase has to be 
 $\sim10^{10}$ cm  with unacceptably large merger time $\sim6\times10^4$ yr. 
 Thus, the model of the merger in the common envelope cannot 
 explain synchronization of violent mass loss and supernova 
 explosion within several hundred years. Therefore, the 
 merger scenario for  SN~2002is and SN~1997cy is unlikely.

The model of the white dwarf explosion in the system 
with a red supergiant component predicts an interesting 
effect. If the hydrogen-rich envelope of the giant is not 
lost completely prior to the explosion, several tenths 
of solar mass will then be lost due to the interaction of the supernova 
envelope and red supergiant. The major fraction of this matter 
acquires low velocity ($\sim 10^3$ km s$^{-1}$) and thus should 
reside in the inner part of the expanding supernova envelope 
(Chugai 1986; Livne et al. 1992). Detection of a narrow H$\alpha$-line 
from the ``inner'' hydrogen in the spectrum of a SN\,Ia 
would indicate an explosion in a system with red supergiant. 
Of course, the presence of a narrow H$\alpha$-line from circumstellar 
gas strongly reduces possibility of the detection of H$\alpha$
from the central region of  SN\,Ia.

\subsection{Single star}
\label{subsec:single}

After pioneering work of Becker and Iben (1979), 
it is usually  supposed 
 that single stars do not form exploding CO cores of 
 Chandrasekhar mass. For instance, Gil-Pons et al. (2003) 
 find that stars with initial mass (8.7 -- 11)$M_{\odot}$ ignite 
 carbon in the weakly degenerate cores and the 
 burning extends to the 
 formation of Ne and Mg. The envelopes of stars are lost by stellar 
 (super)wind. However, uncertainties of the models 
 of the intermediate mass stars do not rule out completely 
  the possibility of formation of exploding CO 
 cores of the Chandrasekhar mass in these stars (Iben and Renzini 1983). 
The resulting supernovae were dubbed SN\,1.5; this name has to 
 stress their intermediate status between   SN\,Ia and  SN\,II.    

One of the uncertainties in the evolutionary computations of the 
limiting mass of CO cores is related to the usually ignored rotation. 
Dominguez et al. (1996)
 approximately took into account the rotation 
in the one-dimensional model and demonstrated that the 
lifting effect permits a star 
with the initial mass of 6.5~$M_\odot$ to form a Chandrasekhar mass 
CO core.

Another obstacle to the SN\,1.5-type supernovae is 
probably a complete loss of the envelope via superwind 
before the mass of the core mounts to $M_{\rm Ch}$.
Remarkably, in this regard,  SN~1997cy exploded in a 
dwarf galaxy (Turatto et al. 2000), likewise, supposedly, SN~2000ic 
(Hamuy et al. 2003). The common property of dwarf galaxies is 
low metallicity that favors low mass loss rate. This 
might explain why the envelopes of presupernovae 
have not been lost by the AGB superwind prematurely.

Let the combination of low metallicity and rotation eventually result 
in formation of a Chandrasekhar mass CO core in a single star. 
What may then cause an  intense mass loss by supergiant in 
$\sim 600$\,yr prior to explosion? One may assume that the 
contraction of the core as it mounts to the Chandrasekhar  mass 
can lead to the intense mass loss. For instance, the increase of 
gravitational potential and, as a consequence, the growth of 
temperature and density may result in a more intense energy release 
in the double burning shell at the core boundary. Another 
possibility may be related to C-burning flashes in the CO core.

The absence of smooth continuum in the early 
spectra of SN~2002ic implies that the mass of the hydrogen in the 
envelope of supernova is significantly smaller than $1~M_\odot$.  
The scenario of single star explosion thus suggests 
that the hydrogen envelope should be almost completely lost 
 prior to the explosion.

\section{Conclusion}
\label{sec:concl}

The simulations of the light curves of SN~2002ic and SN~1997cy 
in the model of the SN\,Ia expansion in a dense circumstellar 
envelope led us to conclusion that
(i) the density of CS 
envelope for both supernovae is comparable and the corresponding 
mass loss prior to SN explosion occurred in the time scale of several 
hundred years with the rate of $\sim10^{-2}~M_{\odot}$ yr$^{-1}$;
(ii) the mass of the CS envelope of SN~1997cy is close to 
$5\,M_\odot$; (iii) the time scale of formation of CS
envelope of SN~1997cy did not exceed 600 yr.

If SN~2002ic and SN~1997cy are actually type Ia supernovae that 
exploded in dense circumstellar envelopes, then there must 
exist a mechanism that provides the synchronization of intense 
outflow of a huge amount of matter in several hundred years 
and the SN~Ia explosion event.
The analysis of two scenarios of the possible evolution to the 
 explosions 
of  SN~2002ic and SN~1997cy  -- accretion onto a white dwarf in 
a symbiotic system or SN\,1.5 scenario -- does not allow us to identify 
with confidence the mechanism of the synchronization of the mass loss 
and explosion. 
We believe that the most natural mechanism should involve 
the contraction of the CO white dwarf (in the 
binary-star scenario) or 
CO core (in the SN~1.5 scenario) when approaching 
the Chandrasekhar
limit. However, the details of the process that 
has to ``switch-on'' the violent mass loss by a supergiant several 
hundred years prior to the SN\,Ia explosion have yet to be understood.
We, however, may almost certainly rule out prima facie 
conceivable scenario of 
the merger  of CO white dwarf and CO core of a red supergiant 
due to  the angular momentum loss via gravitational 
 wave radiation after ejection of the common envelope. 

Both scenarios have interesting predictions that may 
turn out crucial for their verification. The scenario of the
symbiotic star predicts a wide range for the masses of CS
envelopes -- from several tenth of $M_\odot$ to about $6\,M_\odot$, 
with low masses being dominant since low-mass 
giants prevail in symbiotic systems. In the single star 
scenario the initial mass is close to 8\,$M_\odot$ and CS envelopes 
should be, therefore, similar and rather massive. 
If future observations will reveal relatively large number of 
SN~2002ic-type events with the mass of the CS envelope 
greater than one solar, then the single-star model should be preferred.
At present, with only two detected  SN~2002ic-like events, 
both indicating high-mass CS envelopes, the scenario of SN~1.5 is 
favored. 

Note, SN~2002ic-subtype events are extremely rare. 
The fact that, despite their luminosity 
exceeds the luminosity of a typical SN\,Ia, only two 
such supernovae were discovered as yet implies that 
their relative occurrence rate among all SNe\,Ia is 
less than 1 per cent.

\bigskip
This work was partially supported by Russian Foundation for 
Basic Research (grants 01-02-16295 and 03-02-16254) and Federal 
Science and Technology Program ``Astronomy''. 

\newpage

{\bf \large References}

\bigskip

\noindent 1. Arnett W.D. Astrophys. Space Sci. (1969). \\
2. Arnett W.D. Astrophys. J. 237, 541 (1980).\\
3. Becker S.A., Iben I. Jr. Astrophys. J. 232, 831 (1979).\\
4. Chevalier R.A. Astrophys. J. 259, 302 (1982)\\
5. Chugai N.N. Sov. Astron. 30, 563 (1986). \\
6. Chugai N.N. Sov. Astron. 36, 63 (1992). \\
7. Chugai N.N. Mon. Not. R. astron. Soc. 326, 1448 (2001).\\
8. Chugai N.N., Danziger I.J. Astron. Lett. (in press) (2003). \\
9.  Dominguez I., Straniero O., Tornambe A. Astrophys. J. 472, 783 (1996).\\
10. Gil-Pons P., Garcia-Berro E., Jose J., Hernanz M., J. W. Truran J.W.\\
 \indent astro-ph/0306197 (2003). \\ 
11. Hachisu I., Kato M., Nomoto K. Astrophys. J. 522, 487 (1999).\\
12. Hamuy M., Phillips M.M., Suntzeff N.B. {\em et al.} Nature (2003).\\
13. H\"{o}flich P., Khokhlov A.M., Wheeler W.C. Astrophys. J. 444, \\
 \indent 831 (1995).\\
14. Iben I. Jr., Renzini A. Annual Rev. Astron. Astrophys. 21, 271 (1983).\\
15. Landau L.D., Lifshitz E.M. {\em The classical theory of fields},
 (Pergamon \\
\indent Press, Oxford, 1971).\\
16. Livne E., Tuchman Y., Wheeler J.C. Astrophys. J. 399, 665 (1992).\\
17. Turatto M., Suzuki T., Mazzali P. {\em et al.}, Astrophys. J.\\
 \indent 534, L57 (2000).\\
18. Whelan J., Iben I. Jr. Astrophys. J. 186, 1007 (1973). \\
19. Wood P.R. {\em Circumstellar Media in the Late Stages of Stellar \\
\indent Evolution}, Ed. R.E.S. Clegg, I.R. Stevens, W.P.S. Meikle \\
\indent (Cambridge Univ. Press, Cambridge, 1994), p. 15.\\

\newpage

\begin{figure}[t]
\centering\includegraphics[scale=0.8,angle=0]{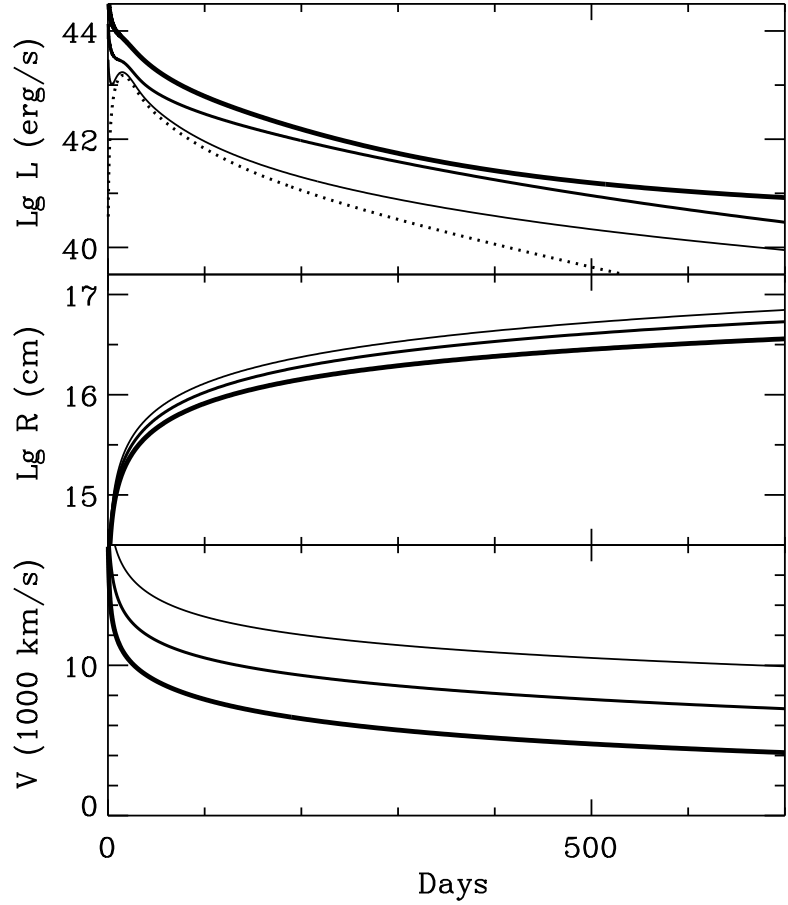}
\vskip 1cm
\caption[]{\footnotesize 
Bolometric light curves (upper panel), radius of the 
thin shell (middle panel), and its velocity
(lower panel) for the model of expansion of SN\,Ia in the stellar wind. 
The light curve for SN\,Ia without wind is shown by dotted line. 
The thickness of lines (from the most thin to the most thick) 
corresponds to the parameter of the wind density $w$ equal to 
$4\times10^{15}$ g cm$^{-1}$,
$2\times10^{16}$ g cm$^{-1}$, and $10^{17}$ g cm$^{-1}$, respectively.    
}
\end{figure} 

\newpage

\begin{figure}[t]
\centering\includegraphics[scale=0.8,angle=0]{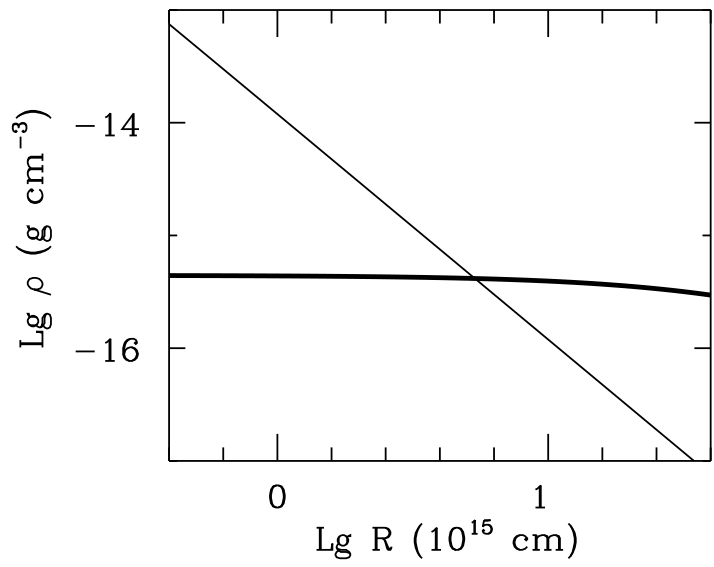}
\caption[]{\footnotesize 
Density distribution in models ic1 (thick line) 
and ic2 (thin line).
}
\end{figure} 

\newpage

\begin{figure}[t]
\centering\includegraphics[scale=0.8,angle=0]{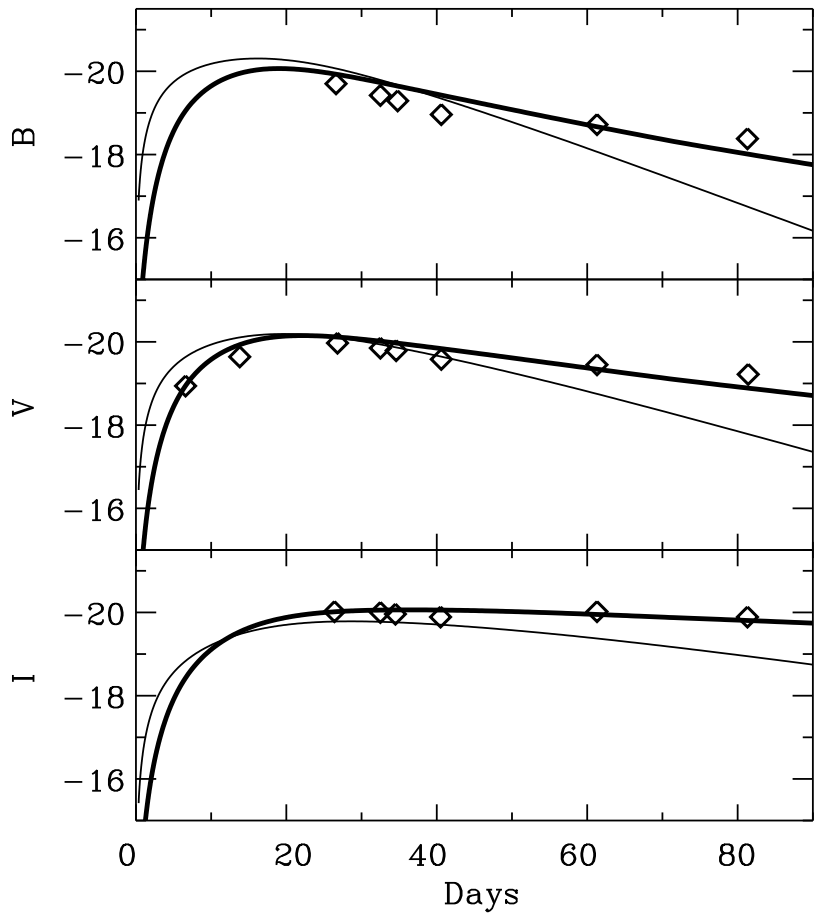}
\vskip 1cm
\caption[]{\footnotesize 
Light curves of SN~2002ic in $BVI$ bands. 
Thick line is the model ic1, thin line is the model ic2. 
Observational data (Hamuy et al. 2003) is shown by diamonds.
}
\end{figure} 

\newpage

\begin{figure}[t]
\centering\includegraphics[scale=0.8,angle=0]{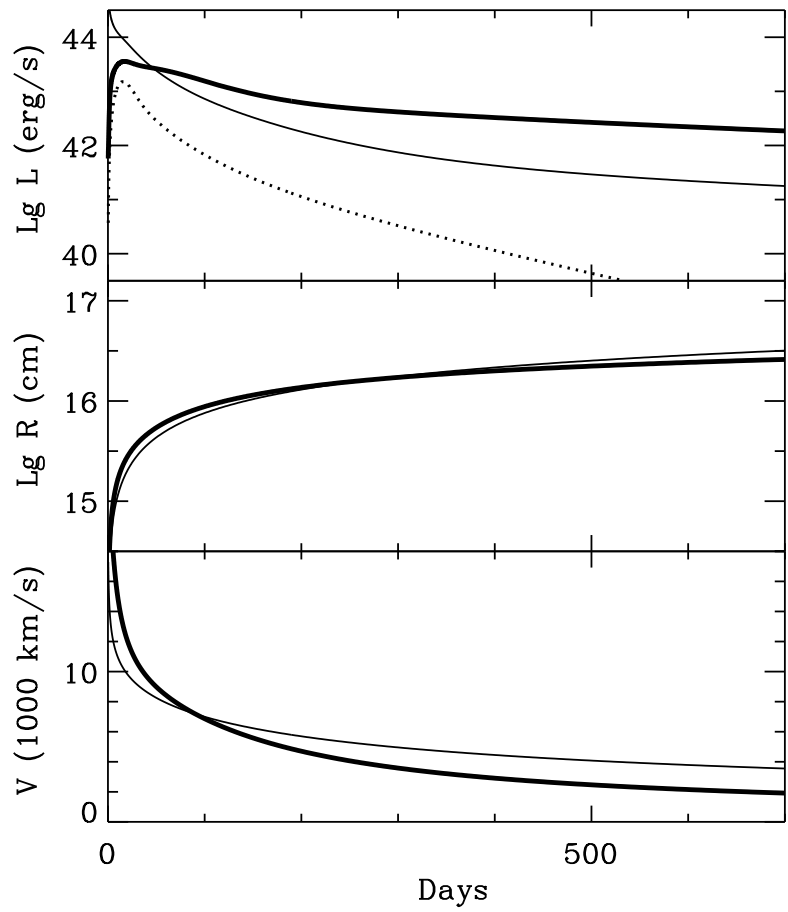}
\vskip 1cm
\caption[]{\footnotesize 
Bolometric light curves, radius, and velocity of the
thin shell for models ic1 (thick line) and ic2 (thin line). 
The dotted line shows the light curve of SN\,Ia in the absence 
of CS gas.
}
\end{figure} 

\newpage

\begin{figure}[t]
\centering\includegraphics[scale=0.8,angle=0]{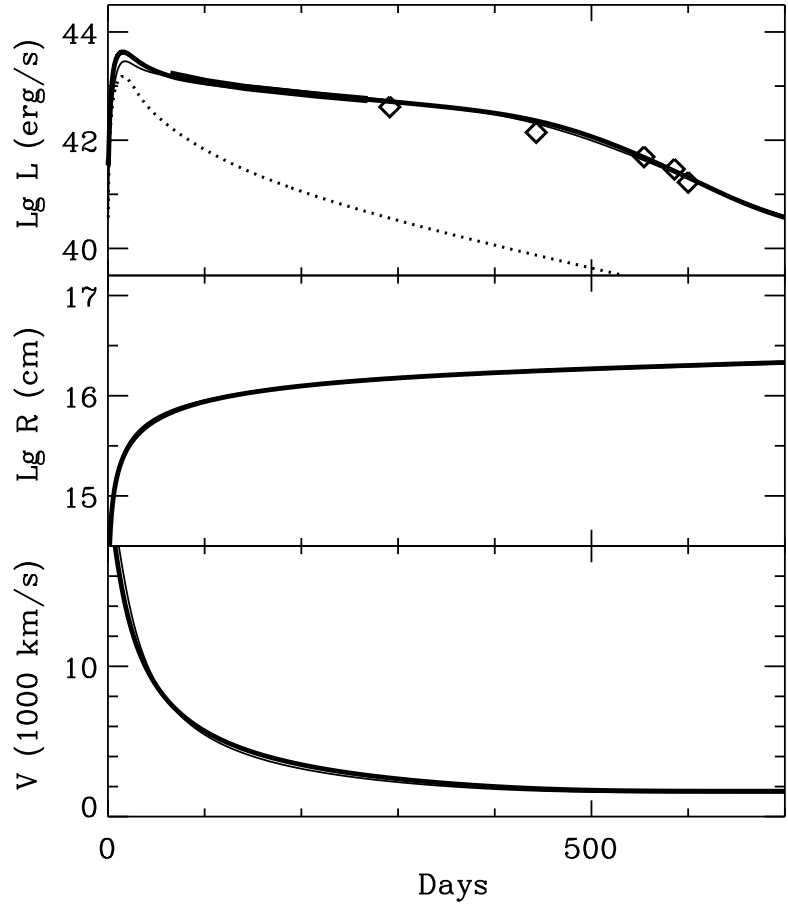}
\vskip 1cm
\caption[]{\footnotesize 
Bolometric light curves, radius, and velocity of the
thin shell for models cy3 (thick line) and cy4 (thin line). 
The dashed curve shows the light curve of SN\,Ia without
CS gas. In the upper panel the diamonds and thick 
line show the empirical light curve (Turatto et al. 2000) 
assuming $H_0 = 70~\mathrm{km~s^{-1}~Mpc^{-1}}$.
}
\end{figure} 

\newpage

\begin{figure}[t]
\centering\includegraphics[scale=0.8,angle=0]{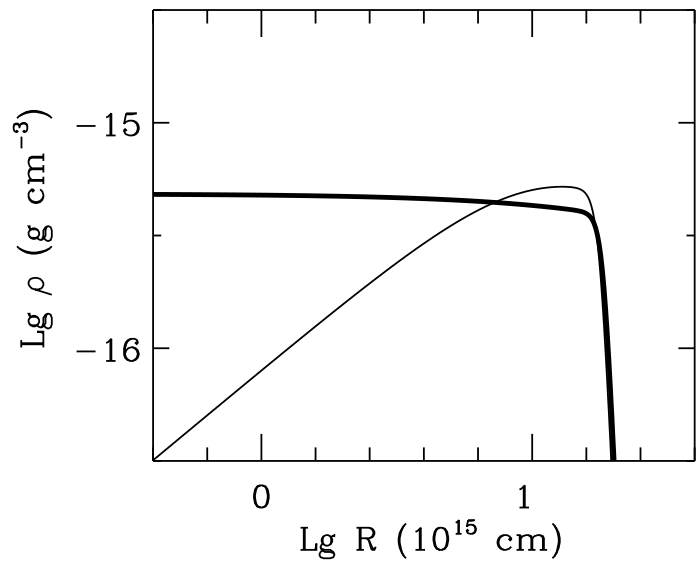}
\caption[]{\footnotesize 
Density distribution in models cy3 
(thick line) and cy4 (thin line).    
}
\end{figure}

\end{document}